\documentclass[aps,pre,twocolumn,superscriptaddress,showpacs]{revtex4}
\usepackage{graphicx}

\begin{document}

\title{Competition between attractive and repulsive interactions in
two-component Bose-Einstein condensates trapped in an optical lattice}
\author{Michal~Matuszewski}
\affiliation{Institute of Theoretical Physics, Physics Department, Warsaw University, Ho%
\.{z}a 69, PL-00-681 Warsaw, Poland}
\author{Boris~A.~Malomed}
\affiliation{Department of Physical Electronics, School of Electrical Engineering,
Faculty of Engineering, Tel Aviv University, Tel Aviv 69978, Israel}
\author{Marek~Trippenbach}
\affiliation{Institute of Theoretical Physics, Physics Department, Warsaw University, Ho%
\.{z}a 69, PL-00-681 Warsaw, Poland}
\affiliation{Soltan Institute for Nuclear Studies, Ho\.{z}a 69, PL-00-681 Warsaw, Poland}

\begin{abstract}
We consider effects of inter-species attraction on two-component gap
solitons (GSs) in the binary BEC with intra-species repulsion, trapped in
the one-dimensional optical lattice (OL). Systematic simulations of the
coupled Gross-Pitaevskii equations (GPEs) corroborate an assumption that,
because the effective mass of GSs is negative, the inter-species attraction
may \emph{split} the two-component soliton. Two critical values, $\kappa _{1}
$ and $\kappa _{2}$, of the OL strength ($\kappa $) are identified.
Two-species GSs with fully overlapping wave functions are stable in strong
lattices ($\kappa >\kappa _{1}$). In an intermediate region, $\kappa
_{1}>\kappa >\kappa _{2}$, the soliton splits into a double-humped state
with separated components. Finally, in weak lattices ($\kappa <\kappa _{2}$%
), the splitting generates a pair of freely moving single-species GSs. We
present and explain the dependence of $\kappa _{1}$ and $\kappa _{2}$ on
thenumber of atoms (total norm), and on the relative strength of the
competing inter-species attraction and intra-species repulsion. The
splitting of asymmetric solitons, with unequal norms of the two species, is
briefly considered too. It is found and explained that the splitting
threshold grows with the increase of the asymmetry.
\end{abstract}

\maketitle

\section{Introduction and the model}

Self-supporting localized patterns in Bose-Einstein condensates (BECs),
which are frequently called matter-wave solitons, have been the subject of
many theoretical and experimental works. The solitons were first
experimentally created in the condensate of $^{7}$Li atoms loaded in a
strongly elongated (\textquotedblleft cigar-shaped") trap \cite{Li}. The use
of the Feshbach-resonance (FR)\ technique \cite{Feshbach} makes it possible
to tune the scattering length of the inter-atomic collisions in the
condensate to a small negative value, thus providing for the weak
self-attractive nonlinearity, which is necessary for the creation of stable
solitons. In another experiment, the solitons were observed in the $^{85}$Rb
BEC remaining in the trap after the onset of collapse induced by the switch
of the interaction between atoms from repulsive to attractive \cite{Wieman}.
Potentially, the FR technique in combination with a quasi-one-dimensional
optical lattice (OL, i.e., a spatially-periodic potential induced by a pair
of counter-propagating laser beams) can be used to create solitons with a
fully three-dimensional shape \cite{ourpaper}.

In most experiments, BEC is created in ultracold gases with repulsion
between atoms. In this case, bright solitons were predicted as a result of
the balance between the repulsive nonlinearity and negative effective mass
of the matter waves, induced by the OL \cite{GS}. The corresponding \textit{%
gap solitons} (GSs) emerge in finite bandgaps of the OL's linear
spectrum, where the negative effective mass is available.
Theoretical models of matter-wave GSs were reviewed in Ref.
\cite{Konotop}, and their stability was analyzed in detail in
Refs. \cite{Pelinovsky} Additionally, in Ref. \cite{Ilya}, the
stability of the GS was also studied against quantum fluctuations,
following the lines of the approach which was earlier developed
for ordinary matter-wave solitons in Ref. \cite{Ami}.
Experimentally, GSs were created in the $^{87}$Rb condensate
trapped in a cigar-shaped potential combined with an OL applied in
the axial direction \cite{Oberthaler}, with $\sim $ 250 atoms in
the established soliton. An essential ingredient of the experiment
was the acceleration of the condensate, with the aim to push the
atomic waves into the spectral region featuring the negative
effective mass. Another approach to the creation of the GS was
proposed in Ref. \cite{Michal}: one may add a strong parabolic
trap to the OL potential, confining the entire condensate to a
small spatial region, and then gradually relax the extra trap,
which may allow the atomic cloud to remain in a relatively compact
GS state. Besides that, one may expect that a chain of GSs may
develop from the modulational instability of nonlinear
quasi-periodic Bloch waves trapped in the OL \cite{GS}.

Binary mixtures of BECs are also available as media in which various
matter-wave patterns may be created. Most typically, the mixtures are formed
by two different hyperfine states of the same atom, such as $^{87}$Rb \cite%
{Rb} and $^{23}$Na \cite{Na}. Creation of a \textit{heteronuclear} mixture
of $^{41}$K and $^{87}$Rb was also reported \cite{hetero}. As concerns the
effective nonlinearity in the mixture, it is known that the sign and
magnitude of the scattering lengths which characterize collisions between
atoms belonging to different species may also be controlled by means of the
FR technique \cite{inter-Feshbach}.

In view of the latter possibility, it is reasonable to consider a binary BEC
with the (natural) intra-species self-repulsion, while the inter-species
interaction is switched to attraction. In recent theoretical works \cite%
{symbiotic}, it was proposed to use this setting to create
\textit{symbiotic} bright solitons: while self-repulsive species
cannot support isolated solitons, the inter-species attraction may
help to create two-component solitons. Moreover, a similar
perspective was discussed in the context of Bose-Fermi mixtures,
where the interaction between bosons is repulsive, but the bosons
and fermions attract each other \cite{Sadhan-BFsoliton}.
Similarly, solitons in a binary degenerate Fermi gas, supported by
the attraction between the fermion species, were predicted
\cite{Sadhan-FFsoliton}. It was also proposed to use the
attraction between fermions and bosons for making bosonic quantum
dots that can trap fermion atoms (in particular, gap solitons in
the BEC trapped in the OL may play the role of such dots)
\cite{Salerno}.

As mentioned above, a salient property of GSs is the negative effective
mass, which should make their dynamical behavior drastically different from
that of ordinary quasi-particles. In particular, both one- \cite{Sakaguchi1}
and two-dimensional \cite{Sakaguchi2} GSs are expelled by the usual
parabolic trapping potential, while being retained by the anti-trapping
(inverted) potential. The negative effective mass may essentially affect the
stability of two-component GSs. If the BEC species repel each other, the
repulsive force, acting in combination with the negative effective mass, may
actually keep the two components together. In Ref. \cite{Arik}, this
possibility was verified, for two- and one-dimensional GSs in the model of
the binary BEC. It was demonstrated that, even with zero intra-species
interaction, the repulsion between the components was sufficient to generate
a family of \textit{symbiotic gap solitons}. Adding nonzero intra-species
repulsion expands the stability region of such GSs. Note that the symbiotic
GSs found in Ref. \cite{Arik} could be both of intra-gap and inter-gap
types, i.e., with the chemical potentials of the two components belonging to
the same or different finite bandgaps in the linear spectrum induced by the
OL.

The objective of the present work is to consider effects of the \emph{%
attraction} between BEC species on two-component GSs trapped in an OL, in
the case when the intra-species interactions are repulsive. As mentioned
above, the necessary signs and magnitudes of the respective nonlinear
coefficients may be adjusted by means of the FR technique. Because of the
negative sign of the effective GS mass, one may expect that the
inter-species attraction \emph{destabilizes} the two-component GSs, with a
trend to split them into separate single-species solitons. This expectation
is confirmed below, by means of systematic simulations of the coupled
Gross-Pitaevskii equations (GPEs) for the macroscopic wave functions of the
two components, $\psi $ \ and $\phi $.

In the normalized form (we set atomic mass, Planck's constant, and the
overall nonlinearity coefficient equal to $1$, and the OL period equal to $%
\pi $), the coupled GPEs take the well-known form \cite{Pethick},%
\begin{eqnarray}
i\psi _{t}+(1/2)\psi _{xx}-\left[ \left( \cos \theta \right) |\psi
|^{2}+\left( \sin \theta \right) |\phi |^{2}\right] \psi  \nonumber \\
+\kappa \cos (2x)\psi =0,  \label{eq1}
\end{eqnarray}
\begin{eqnarray}
i\phi _{t}+(1/2)\phi _{xx}-\left[ \left( \cos \theta \right) |\phi
|^{2}+\left( \sin \theta \right) |\psi |^{2}\right] \phi  \nonumber \\
+\kappa \cos (2x)\phi =0,  \label{eq2}
\end{eqnarray}
where $\kappa $ is the strength of the OL (actually measured in units of the
corresponding recoil energy), and angle $\theta $ is a parameter that
determines the relative strength and sign of the inter- and intra-species
interactions. The case of the intra-species repulsion and inter-species
attraction, which we aim to consider in this work, corresponds to $-\pi
/2<\theta <0$. If $\psi $ and $\phi $ represent two components of a spinor
BEC (therefore, equal atomic masses are assumed in the two equations) with
opposite $z$-components of the hyperfine spin, $m_{F}=\pm 1$, then $\tan
\theta =\left( a_{0}+a_{2}\right) /\left( a_{0}-a_{2}\right) $, with
coefficients $a_{0}$ and $a_{0}$ accounting for the mean-field
(spin-independent) and spin-exchange interactions between the atoms \cite{Ho}%
. These coefficient may be, if necessary, controlled by means of the FR
technique, as mentioned above.

Equations (\ref{eq1}) and (\ref{eq2}) conserve the corresponding
Hamiltonian, and two norms (scaled numbers of atoms in the different
species), $N_{\Psi ,\Phi }\equiv \int_{-\infty }^{+\infty }\left\{
\left\vert \psi (x)\right\vert ^{2},\left\vert \phi (x)\right\vert
^{2}\right\} dx$. We will chiefly consider \textit{symmetric} GS complexes,
with $N_{\Psi }=N_{\Phi }\equiv N$ (Section II); asymmetric states with
unequal norms will also be considered, but briefly, in Section III.

\section{Results: symmetric gap solitons}

Stationary soliton solutions to Eqs. (\ref{eq1}) and (\ref{eq2}) are looked
for as
\begin{equation}
\psi (x,t),\phi (x,t)=e^{-i\mu t}\Psi (x),e^{-i\nu t}\Phi (x),  \label{stat}
\end{equation}%
where $\mu $ and $\nu $ are chemical potentials of the two components. In
this work, we focus on the two-component GSs of the most fundamental type,
with both $\mu $ and $\nu $ falling in the first finite bandgap of the
linear spectrum.

The starting point is the symmetric soliton with identical components, $\mu
=\nu $ and $\Psi (x)=\Phi (x)$, where the real wave function, $\Psi (x)$,
obeys the ordinary stationary equation for the single-component BEC,
\begin{equation}
\mu \Psi +(1/2)\Psi ^{\prime \prime }-\sqrt{2}\left( \sin \left( \theta +\pi
/4\right) \right) \Psi ^{3}+\kappa \cos (2x)\Psi =0.  \label{U}
\end{equation}%
This equation has GS solutions provided that $\sin \left( \theta +\pi
/4\right) >0$, which actually implies $-\pi /4<\theta <0$, since we are
interested in $\theta <0$ (the inter-species attraction), as said above. The
remaining interval, $-\pi /2\leq \theta <-\pi /4$, corresponds to the
ordinary symbiotic solitons \cite{symbiotic} trapped in the OL, with $\mu $
falling in the semi-infinite gap.

\begin{figure}[tbp]
\includegraphics[width=8.5cm]{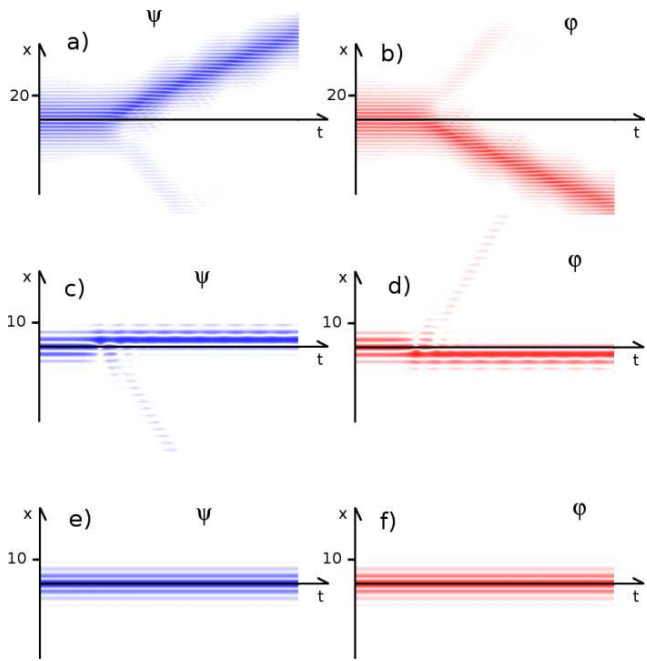}
\caption{(Color online) Three generic scenarios of the evolution of
two-component symmetric gap solitons are shown by means of density-level
contours of the two species in the $\left( x,t\right) $ plane. a,b)
Splitting into freely moving solitons in a weak lattice, $\protect\kappa =0.2
$. c,d) Formation of a stable bound states of separated solitons in a
moderately strong lattice, $\protect\kappa =0.4$. e,f) Stability of the
bound state in a strong lattice, $\protect\kappa =0.6$. Other parameters are
fixed: $\protect\theta =-0.3$, norm of each component $N=1.5$, and the
evolution time, $t=500$.}
\label{fig1}
\end{figure}

Equation (\ref{U}) can be solved by means of known numerical methods \cite%
{GS,Sakaguchi1} (the variational approximation \cite{Progress}, which was
first used in the framework of the GPE in Ref. \cite{firstVA}, becomes quite
cumbersome if applied to GSs \cite{OurPaper}, therefore we do not resort to
this method here). We tested the stability of symmetric solitons, generated
by Eq. (\ref{U}), against small random perturbations (which include a small
initial separation between the components) by means of direct simulations of
Eqs. (\ref{eq1}) and (\ref{eq2}). In agreement with the expectation that the
effective negative mass of the GS can make the two-component bound state
unstable against the splitting under the action of the inter-species
attraction, we have observed three different scenarios of the
perturbation-induced dynamics, depending on the OL strength, $\kappa $. As
shown in Figs. \ref{fig1} a) and c), the bound state splits into two freely
moving single-species solitons in the weak lattice, and remains stable in
the strong OL. In the intermediate case, Fig. \ref{fig1} b), the bound
states also splits, but the components cannot move freely; instead, they get
pinned at a finite distance between them. As concerns the free motion of the
single-component soliton, in Ref. \cite{Sakaguchi1} it was demonstrated that
the GS moves without any tangible loss if its amplitude does not exceed a
certain maximum value, above which the moving soliton is being braked by the
underlying lattice.

\begin{figure}[tbp]
\includegraphics[width=8.5cm]{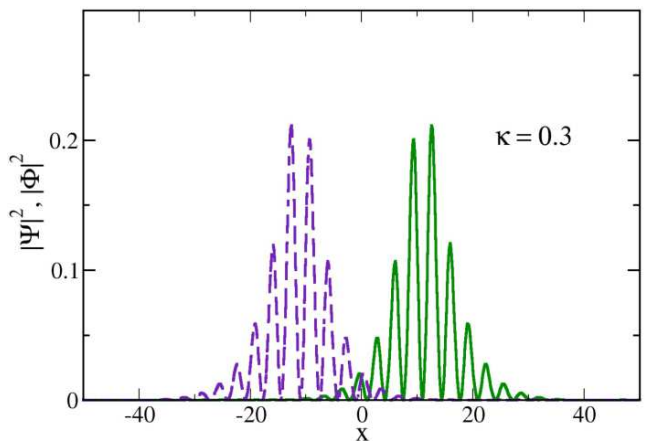}
\caption{(Color online) A stable bound state of two gap solitons formed in a
moderately strong lattice as a result of the splitting of the unstable state
with overlapping wave functions, cf. Fig.~\protect\ref{fig1} b) (parameters
are as in that figure, except that $\protect\kappa =0.3$).}
\label{fig2}
\end{figure}

\begin{figure}[tbp]
\includegraphics[width=8.5cm]{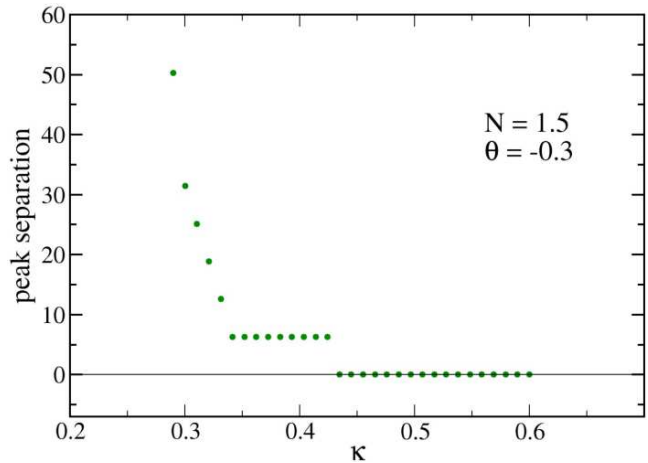}
\caption{(Color online) The separation between the peaks of the $\Psi $- and
$\Phi $-components in the stable two-component soliton, generated by
splitting of the unstable state with overlapping wave functions [see Fig.
\protect\ref{fig1} b)], versus the lattice strength.}
\label{fig4}
\end{figure}

A typical example of the stable symmetric soliton with separated centers of
its two components is displayed in Fig. \ref{fig2}. In addition, the
separation between centers of the two species in the stable solitons of this
type, generated by the splitting of the unstable soliton with overlapping
wave functions, $\Psi =\Phi $, is shown in \ref{fig4}. According to Fig. \ref%
{fig1}, the separation is zero at $\kappa >0.46$, as the overlapping bound
state is stable in that case, while at $\kappa <0.29$ the split components
do not come to a halt (i.e., the separation is infinite).

\begin{figure}[tbp]
\includegraphics[width=8.5cm]{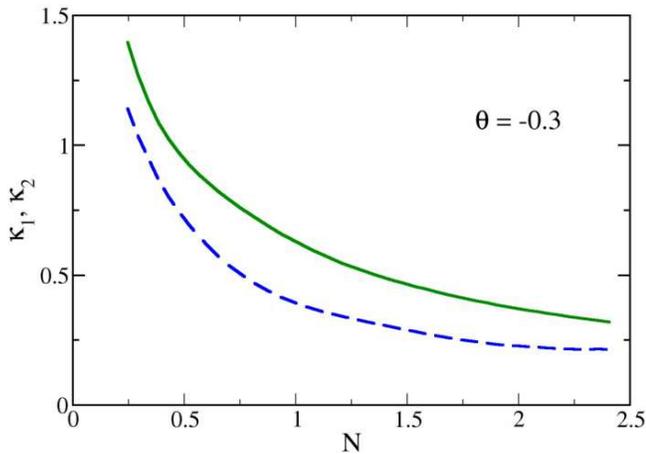}
\caption{(Color online) Critical values of the lattice depth, $\protect%
\kappa _{1}$ (solid line), below which the solitons with the overlapping
wave functions, $\Psi =\Phi $, start to split, and $\protect\kappa _{2}$
(dashed line), below which the single-component solitons generated by the
splitting move freely, are shown as functions of norm $N$ of each component.}
\label{fig5}
\end{figure}

As suggested by Fig. \ref{fig4}, two critical values, $\kappa _{1}$ and $%
\kappa _{2}$, of OL strength $\kappa $ may be identified, for given $\theta $
and norm $N$: at $\kappa <\kappa _{1}$, the original soliton, with $\Psi
(x)=\Phi (x)$, is unstable and splits, and at $\kappa <\kappa _{2}$ (with $%
\kappa _{2}<\kappa _{1}$) the splinters (single-component GSs) are not
pinned by the lattice, but rather move freely. The dependence of both
critical values on the norm of the initial overlapping bound state is shown
in Fig. \ref{fig5} (to interpret the results in physical terms, recall that $%
\kappa $ shown in these plots is measured in units of the recoil energy).
The decrease of $\kappa _{1}$ and $\kappa _{2}$ with increase of $N$ is easy
to understand, because the solitons with larger $N$ are narrower, hence they
are stronger pinned by the lattice \cite{Sakaguchi1}.

\begin{figure}[tbp]
\includegraphics[width=8.5cm]{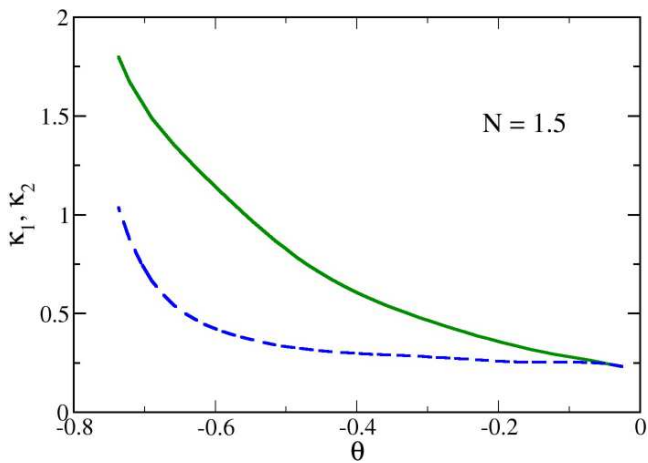}
\caption{(Color online) The critical lattice strengths, $\protect\kappa _{1}$
(solid) and $\protect\kappa _{2}$ (dashed), as functions of angle $\protect%
\theta $, which determines the relative strength of the inter-species
attraction and intra-species repulsion.}
\label{fig6}
\end{figure}

In addition, Fig. \ref{fig6} shows the dependence of $\kappa _{1}$ and $%
\kappa _{2}$ on $\theta ,$ at fixed $N$. The increase of the critical values
of the lattice strength with the growth of $\left\vert \theta \right\vert $
is quite natural: as the inter-species attraction, which actually pushes the
two GS components apart, gets stronger with respect to the intra-species
repulsion, which tends to stabilize the GS, it becomes more difficult for
the lattice to hold the components together. Both critical values diverge in
the limit of $\left\vert \theta \right\vert \rightarrow \pi /4$, since point
$\theta =-\pi /4$ is a singular one, as follows from Eq. (\ref{U}). For
values of $\theta $ close to zero, the threshold $\kappa _{2}$ is almost
constant, because it is mainly determined by the ability of the lattice to
stop the motion of the single-component soliton after the splitting.

\section{Asymmetric gap solitons}

\begin{figure}[tbp]
\includegraphics[width=8.5cm]{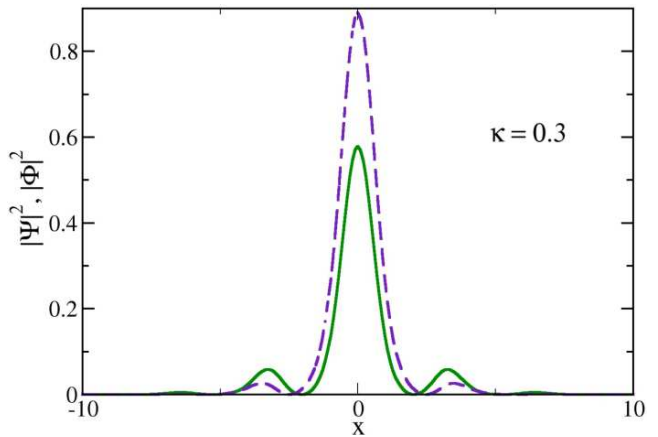}
\caption{(Color online) A stable unsplit soliton, with unequal norms in the
two components: $N_{\Psi }=1.5$, $N_{\Phi }=1.6$, and $\protect\theta =-0.3$%
, $\protect\kappa =1$.}
\label{fig8}
\end{figure}

We have also investigated the dynamics of asymmetric solitons, with
different norms of the two components, $N_{\Psi }\neq N_{\Phi }$. An example
of a stable soliton of that type, with coinciding centers of its components,
is displayed in Fig. \ref{fig8}.

\begin{figure}[tbp]
\includegraphics[width=8.5cm]{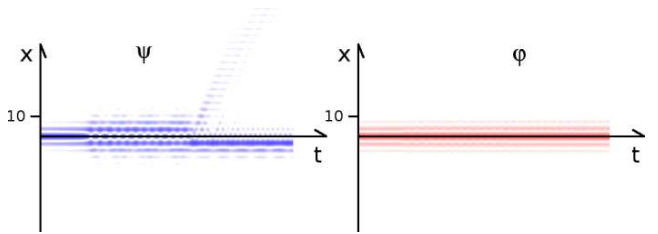}
\caption{(Color online) Spontaneous splitting of an asymmetric soliton below
the threshold, $\protect\kappa _{1}$, for $N_{\Psi }=1.5$, $N_{\Phi }=1.6$, $%
\protect\theta =-0.3$, and $\protect\kappa =0.5$ (in this case, $\protect%
\kappa _{1}$ is slightly larger than $0.5$, see Fig. \protect\ref{fig10}).
Total evolution time is $t=1000$.}
\label{fig9}
\end{figure}

Similarly to the symmetric case, the asymmetric solitons become unstable
against the splitting of the two components if the lattice strength falls
below the threshold value, $\kappa _{1}$, see an example in Fig. \ref{fig9}.
In the simulations, the splitting follows onset of oscillations of the
solitons, with a significant amplitude.

\begin{figure}[tbp]
\includegraphics[width=8.5cm]{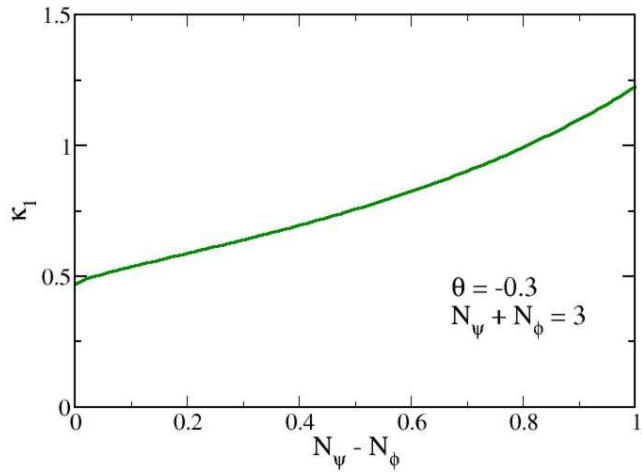}
\caption{(Color online) The dependence of the lattice-strength splitting
threshold, $\protect\kappa _{1}$, for asymmetric solitons on the norm
difference between the components (i.e., the asymmetry degree).}
\label{fig10}
\end{figure}

Naturally, $\kappa _{1}$ depends on the asymmetry parameter, $\left( N_{\Psi
}-N_{\Phi }\right) /\left( N_{\Psi }+N_{\Phi }\right) $, as shown in Fig. %
\ref{fig10}. The growth of the splitting threshold with the increase of the
asymmetry can be readily explained. Indeed, as the norm of component $\Phi $
becomes smaller, its width increases. On the other hand, the effective
pinning force (the amplitude of the effective Peierls-Nabarro potential \cite%
{PN}) acting on the soliton exponentially decays with the increase of the
soliton's width, hence a higher strength of the OL is required to prevent
the splitting instability of the two-component asymmetric soliton.

\section{Conclusions}

We have considered effects of the attraction between two species in the
binary BEC with intra-species repulsion, trapped in the OL (optical
lattice), on the two-component GSs (gap solitons), supported by the balance
between the repulsion and negative effective mass induced by the OL
potential. Systematic simulations confirm the prediction suggested by the
fact that the effective mass of the gap soliton is negative: if the OL is
not strong enough, the inter-species \emph{attraction} results in \emph{%
splitting} of the two-component GS. We have identified two threshold values,
$\kappa _{1}$ and $\kappa _{2}$, of the OL strength ($\kappa $), for the
two-component GS with equal norms of its components. The unsplit solitons
(with fully overlapping wave functions, $\Psi =\Phi $) are stable in strong
lattices, with $\kappa >\kappa _{1}$; they split into a stationary symmetric
state with separated components in interval $\kappa _{1}>\kappa >\kappa _{2}$
, and, in weak lattices, with $\kappa <\kappa _{2}$, the splitting generates
a pair of freely traveling single-species GSs. The dependences of $\kappa
_{1}$ and $\kappa _{2}$ on the norm of the original overlapping soliton, and
on the relative strength of the inter-species attraction and intra-species
repulsion ($\theta $) were found and explained. We have also considered, in
a brief form, the dynamics of asymmetric solitons, with unequal norms of the
two species. In particular, it was found (and explained) that the splitting
threshold, $\kappa _{1}$, grows with the increase of the relative asymmetry.

The model introduced in this work calls for further analysis. In particular,
it may be interesting to explore effects of \emph{intra-species attraction}
on symbiotic GSs supported by the \emph{inter-species repulsion} (such as
two-component GSs found in Ref. \cite{Arik}), i.e., the case exactly
opposite to that considered above. Another relevant generalization may be to
study similar effects in two-dimensional GSs build of two species.

\section{Acknowledgements}

M.M. acknowledges support from the Foundation for Polish Science. M.T. was
supported by the Polish Ministry of Scientific Research and Information
Technology under grant N202 022 32/0701. B.A.M. appreciates hospitality of
the Institute of Theoretical Physics and Soltan Institute for Nuclear
Studies at the Warsaw University (Warsaw, Poland). The work of this author
was supported, in a part, by the Israel Science Foundation through
Excellence-Center grant No. 8006/03, and by the German-Israel Foundation
through grant No. 149/2006.

\end{document}